\documentclass[useAMS,usegraphicx,usenatbib,twocolumn]{mn2e}
\usepackage{amssymb}

\def\aj{AJ}% Astronomical Journal

\def\apj{ApJ}% Astrophysical Journal
\def\apjl{ApJ}% Astrophysical Journal, Letters
\def\apjs{ApJS}% Astrophysical Journal, Supplement
\def\aap{A\&A}% Astronomy and Astrophysics
\def\mnras{MNRAS}% Monthly Notices of the RAS
\def\nat{Nature}% Nature
\def\aaps{A\&AS}
\def\apss{APSS}
\def\fcp{FCP}

\title[Bars and Interactions]{A numerical study of interactions and stellar bars}

\author[Martinez-Valpuesta et al.]{\parbox{\textwidth}{Inma Martinez-Valpuesta,$^{1,2}$\thanks{E-mail:
imv @ iac.es}, J. Alfonso L. Aguerri,$^{1,2}$ A. C\'esar Gonz\'alez-Garc\'{\i}a,$^{3}$\\
and Claudio Dalla Vecchia$^{1,2}$ and Martin Stringer$^{1,2}$}\vspace{0.6cm}\\
$^{1}$Instituto de Astrof\'isica de Canarias, E-38205 La Laguna, Tenerife, Spain\\
$^{2}$Universidad de La Laguna, Dpto. Astrof\'isica, E-38206 La Laguna, Tenerife, Spain\\
$^{3}$Instituto de Ciencias del Patrimonio, CSIC, Santiago de Compostela, 15704, A Coru\~na, Spain
}
\begin{document}
\date{Accepted xxxx December xx. Received xxx December xxx; in original form
xxx}

\pagerange{\pageref{firstpage}--\pageref{lastpage}} \pubyear{2012}

\maketitle

\label{firstpage}

\begin{abstract}
For several decades it has been known that stellar bars in disc galaxies can be triggered by interactions, or by internal processes such as dynamical instabilities. In this work, we explore the differences between these two mechanisms using numerical simulations. We perform two groups of simulations based on isolated galaxies, one group in which a bar develops naturally, and another group in which the bar could not develop in isolation. The rest of the simulations recreate 1:1 coplanar fly-by interactions computed with the impulse approximation. The orbits we use for the interactions represent the fly-bys in groups or clusters of different masses accordingly to the velocity of the encounter. In the analysis we focus on bars' amplitude, size, pattern speed and their rotation parameter, ${\cal R}=R_{CR}/R_{bar}$. The latter is used to define fast (${\cal R}<1.4$) and slow rotation (${\cal R}>1.4$). Compared with equivalent isolated galaxies we find that bars affected or triggered by interactions: (i) remain in the slow regime for longer; (ii) are more boxy in face-on views; (iii) they host kinematically hotter discs. Within this set of simulations we do not see strong differences between retrograde or prograde fly-bys. We also show that slow interactions can trigger bar formation.
\end{abstract}


\section{Introduction}
\label{sec:Intro}

Bars are ellipsoidal-like features present in a large fraction of discs galaxies. About 40-50 $\%$ of the galactic discs in the local Universe observed in the optical show a bar structure \citep[see e.g.,][]{Marinova+07, Aguerri+09, Diaz-Garcia+15}. This fraction increases up to $\sim 60-70\%$ when observing in the near-infrared \citep[see][]{Eskridge+00,Menendez-Delmestre+07}. The fraction of barred galaxies depends on several integrated properties such as: galaxy stellar mass, star formation history, and colour \citep[see][]{Nair+10, Masters+11, Mendez-Abreu+12}. The fraction decreasing with increasing redshift \citep[see][]{Melvin+14}, though this variation depends strongly on other internal galaxy properties such as mass, colour or bulge prominence \citep[see e.g.,][]{Sheth+08}.

Bars have a strong influence on the dynamics of disc galaxies. In particular, the presence of bars highly influences the exchange of angular momentum between the different galaxy components, mainly halo and disc, \citep[see][]{Lynden-Bell+72, Tremaine+84, Weinberg+85, Debattista+98, Debattista+00, Athanassoula02, Martinez-Valpuesta+06, Saha+12}. They are also related to gas inflow and star formation events \citep[see e.g.,][]{Hernquist+95, Martinet+Friedli97, Aguerri99}. The growth and feeding of central supermassive black holes in galaxies can also be driven by bars \citep[see][]{Shlosman+90, Corsini+03}.

Three main observational parameters characterise bars in galaxies: the length, the strength and the pattern speed. The bar length determines the extension in the disc of the orbits building the bar \citep{Contopoulos81}. The length of bars has been determined observationally with different methods: optical visual inspection \citep[see e.g.,][]{Kormendy79, Martin95}, locating the maximum of the isophotal ellipticity \citep[see][]{Wozniak+95, Marquez+99, Laine+02, Marinova+07, Aguerri+09,Diaz-Garcia+15}, or structural decompositions of the galaxy surface brightness \citep[see][]{Prieto+97, Prieto+01, Aguerri+01, Aguerri+03, Aguerri+05, Laurikainen+09, Gadotti+08, Weinzirl+09}. 

The bar strength measures non-axisymmetric forces produced by the bar potential. It is determined with several methods: measuring the torques of the bar from photometry \citep[see][]{Combes+Sanders81, Quillen+94, buta+Block01, Salo+10, Diaz-Garcia+15}, or from kinematics \citep{Seidel+15}; measuring the bar ellipticity \citep[see][]{Martinet+Friedli97,Aguerri99,Whyte+02}; or with Fourier decomposition of the galaxy light \citep[see][]{Ohta+90, Marquez+96,Aguerri+00,Laurikainen+05,Diaz-Garcia+15}. 

The bar pattern speed is defined as the rotational frequency of the bar. This dynamical parameter has been determined in the literature though several methods, among them: hydrodynamical simulations of barred galaxies \citep[see e.g.,][]{Lindblad+96, Laine+99, Weiner+01, Aguerri+01, Perez+04, Treuthardt+08}, identifying galaxy structures with Lindblad resonances \citep[see][]{Buta+96, Munoz-Tunon+04, Perez+12}, changes in the morphology or phase of spiral arms with radius \citep[see][]{Puerari+97, Aguerri+00, Sierra+14}; the so-called Tremaine-Weinberg method \citep[see][]{Kent+87,Merrifield+95,Gerssen+99, Debattista+02, Aguerri+03, Corsini+03, Debattista+04, Corsini+07, Treuthardt+07, Aguerri+15}; or studying the morphology of the residual gas velocity field after the rotation velocity subtraction \citep{Sempere+95, Font+11, Font+14}.

Observationally, it has been shown that bar parameters depend on morphological galaxy properties. Thus, length and strength depend on the galaxy Hubble type. In particular, S0 galaxies show larger bars than late-type ones \citep[see][]{Elmegreen+95, Aguerri+09, Erwin+05, Menendez-Delmestre+07}, but see also \citet{Masters+11} for a different perspective. In addition, S0 galaxies show weaker bars than late-type ones \citep[see][]{Laurikainen+07, Aguerri+09, Buta+10}. The dependence of the bar pattern speed on the morphological type is not so evident \citep[see][]{Aguerri+15}. 

Numerical simulations show that bar parameters evolve with time in isolated discs. These simulations show that evolution proceeds through three main phases. The first phase corresponds to the bar formation and extends $\sim2$~Gyr. During this period the bar strength and bar length grows rapidly. The bar is formed and clearly visible in the disc. During the second phase the bar buckles \citep{Raha+91,Combes+Sanders81}. The time extension of this phase is about $1$~Gyr, and the bar becomes shorter and weaker \citep{Martinez-Valpuesta+Shlosman04}. The third phase of the bar formation extends for several Gyrs and is generally called secular evolution epoch. During this phase bar grows slowly by increasing its length and strength. In contrast, its pattern speed continuously decreases through all these phases \citep[see][]{Martinez-Valpuesta+06}. The rate at which bar parameters change during the three phases depends on several properties of the galaxy: dark matter content \citep[][]{Debattista+98, Debattista+00}, gas content \citep{Athanassoula+13}, the disc kinematics \citep[see][]{Athanassoula96}, and even the shape of the dark matter halo \citep[see][]{Athanassoula03}. 

The general evolution described above is all based on simulations of isolated galaxies. But we know that galaxies interact. Are there any external influences in the formation and evolution of bars? What is the `nature' vs. `nurture' of bars? Which is the role played by the environment in the bar formation and evolution? There are examples in the Universe of isolated galaxy pairs showing prominent bar features \citep[see e.g.,][]{Fuentes-Carrera+04}. These cases point towards the influence of the environment on the bar formation which is also studied in other galaxy samples. The pioneering work of \cite{Thompson+81} showed an increased fraction of barred galaxies in the central region of the Coma cluster indicating that tidal interactions trigger bar formation. Similar results were also found in other samples, especially for early-type galaxies \citep[see][]{Giuricin+93, Andersen+96, Eskridge+00, Barazza+09, Lansbury+14, Lin+14}. More recently, \cite{Mendez-Abreu+12} showed that the effect of the environment on the bar formation depends on the mass of the galaxy. They proposed that interactions trigger bar formation in massive galaxies with many interactions, are stable enough to keep their cold discs and therefore form bars. In contrast, the discs of low-mass haloes are heated by interactions inhibiting the bar formation.

N-body simulations have shown that interactions trigger bar formation in discs stable against their development in isolation \citep[see][]{Noguchi87, Aguerri+09, Lang+14}. The strengths and the angular velocities of the bars change due to resonant transfer of angular momentum or mass loss from the end of the bar produced by interactions \citep[see e.g.][]{Gerin+90, Sundin+93, Miwa+Noguchi98}. The variation of the bar parameters produced by tidal effects depends on the mass of the perturber and/or the relative phase of the bar and the companion at pericenter \citep[see][]{Gerin+90,Sundin+93}. \cite{Miwa+Noguchi98} found that bars induced by simulations are confined to the Inner Lindblad Resonant (ILR), producing slow bars. 

Simulations including gravity and hydrodynamics have shown that the gas phase plays an important role in bars formed by tidal events. In this case, depending on the orbital parameters of the perturber, interactions tend to speed up the transition from a galaxy with a strong bar to one with a weak or even without a bar \citep[see][]{Berentzen+03}. Moreover, cosmological simulations show that bars can be formed and destroyed several times during a galaxy's life-time depending on the accretion history \citep[see][]{Romano-Diaz+08}. The mass ratio between the main galaxy and the perturber creating the interaction determines bar formation in the main galaxy, in the perturber or even in both \citep[see e.g.][]{Kazantzidis+11, Lang+14, Lokas+14}.

In this work, we revisit the case of the influence of interactions on bar formation. The aim of this paper is to analyse the change produced by interactions on the observable bar parameters. Several high-resolution N-body simulations have been run to achieve this goal. 

This paper is organized as follows. Sect. \ref{sec:Simu} shows the description of the simulations. The main results are given in Sect.~\ref{sec:effects},~\ref{sec:phot} and \ref{sec:kin}. The discussion and conclusions are shown in Sect. \ref{sec:dis} and \ref{sec:con}, respectively.

\section{Description of Simulations}
\label{sec:Simu}

Our main goal is to determine dynamical differences between galactic bars that are self-generated and those that are purely induced by tidal interactions. Our approach uses N-body numerical simulations. The simulations have been run with an improved version of FTM4-4 \citep{Heller+Shlosman94}, using the potential solver falcON \citep{Dehnen02}. We have run two fiducial simulations in isolation. The simulation is initiated with:
$5\times10^5$ particles describing the exponential disc and the same number for the dark matter halo. This number of particles assures that we achieve the necessary resolution to resolve resonances and therefore properly follow the formation and evolution of the bar \citep[see e.g.][]{Dubinski+09,Weinberg+Katz07a,Weinberg+Katz07b}. The halo distribution is generated following \citet{Fall+Efstathiou80}. The rotating exponential disc is set with Toomre $Q=1.5$, scale length of 2.85~kpc. The gravitational softening is 160~pc for all particles and the halos extend out to $\sim 30$~kpc. The only difference between these two simulations is the fraction of baryonic matter. For SIMI0, $30\%$ of the total mass within a radius of $7$~kpc is in the disc. In the case of SIMI1, $50\%$ of the total mass within $7$~kpc is in the disc. The mass of these galaxies is $3.99\times10^{11}$~M$_{\sun}$ for SIMI0 and $2.17\times10^{11}$~M$_{\sun}$ for SIMI1. At the edge of the disk the stars are rotating at $\sim 200$~km/sec. With these settings we have two types of isolated galaxies: one which does not develop a bar in isolation (SIMI0) and one which develops a strong bar (SIMI1). 

We are interested in fast fly-by interactions, where there is insufficient time for the systems to react during the encounter, and all the effects develop after the interaction have taken place \citep{Gonzalez-Garcia+05}. In order to speed up the numerical calculations we have modelled the interaction with the impulse approximation (IA, see below). The modified model is then run in isolation for $\sim 5 $~Gyr in order to see the effects of such interaction. 

\subsection{Impulse approximation}

By applying the impulse approximation, we assume that the galaxy has no time to reorganise during the encounter, so that all the energy is injected as kinetic energy. Reorganisation within the potential of the host halo occurs after the encounter has finished. We also make use of formulas based on the tidal approximation \citep[][\S\,7.2d]{Binney+Tremaine87}, in which the tidal field has been expanded to first order.  

In the tidal approximation, after a fast interaction with a perturber of mass $M_1$, the change of velocity of the stars of the perturbed galaxy, $\Delta V_2$, scales linearly with the galactocentric radius, $R$. The standard tidal approximation considers the perturber as a point mass. Following \citet{Gnedin+99}, \citet{Gonzalez-Garcia+05} generalised the approximation equations in order to take into account not only the extended nature of the perturber, but also the rotation of the perturbed galaxy and the duration of the encounter. From their equation~(4), which gives the variation of kinetic energy, the absolute value of the velocity perturbation is

\begin{equation}
\mid\Delta V_2(R)\mid\sim 2\, \frac{GM_1}{b^2V}\frac{R_{\rm peri}}{R_{\rm max}} R\,(1-\omega\tau)^{-1.25}\,,\end{equation}
where $R_{\rm peri}$ is the pericenter distance, $b$ is the orbital impact parameter, $V$ is the relative velocity at the pericenter passage and $R_{\rm max}=15~\mathrm{kpc}$ the perturber's cut-off radius. The rightmost term includes the rotation frequency of the perturbed galaxy, $\omega$, and the encounter's duration, $\tau$.

All simulations were performed assuming that the encounters are between galaxies of the same size and mass. In addition, the disc of the galaxy and the orbit of the perturber are coplanar. Therefore, encounters are either prograde or retrograde. Initial conditions were generated for prograde and retrograde encounters, the latter by keeping the same orbital configuration and inverting the spin of the perturbed galaxy by flipping the particle distribution. The orbit has pericenter at distance $R_{\rm peri}=30~\mathrm{kpc}$ from the centre of the galaxy. The relative velocity at the pericenter passage is 500, 1000 and $2000~\mathrm{km}\,\mathrm{s}^{-1}$ to ensure that the orbit is hyperbolic, and that we are comfortably in the regime of non-merging initial conditions \citep{Gonzalez-Garcia+VanAlbada05}. The parameters of the simulations are summarised in Table~\ref{tab:main}.

For model SIMI1, which develops a bar and experiences a buckling instability, we would like to see the impact of the interaction at the two different evolutionary stages, before and after the buckling. We introduce the IA at different times (see Table~\ref{tab:main}), one before the buckling ($t11$) and one after the buckling ($t12$) when the bar has already resumed its evolution.

\begin{table}
\begin{tabular}{l|c|c|c|}
\hline\hline
NAME &  $R_{peri}$(kpc) & Spin & Time \\
\hline
I0\_d\_2000  & 30 & P &  $t01=1.45$~Gyr \\
I0\_d\_1000  & & & \\
I0\_d\_500   & & & \\
I0\_i\_2000  & 30 & R & $t01=1.45$~Gyr  \\
I0\_i\_1000  & & & \\
I0\_i\_500   & & & \\
\hline\hline
I1\_d\_2000   & 30 & P &  $t11=1.45$~Gyr \\ 
I1\_d\_1000   & & & \\
I1\_d\_500    & & & \\
I1\_i\_2000   & 30 & R &  $t11=1.45$~Gyr \\
I1\_i\_1000   & & & \\
I1\_i\_500    & & & \\
\hline\hline
I1.00.30.d.t12.h   & 30 & P & $t12=4.80$~Gyr  \\
I1.00.30.i.t12.h   & 30 & R &  $t12=4.80$~Gyr \\
\hline\hline
\end{tabular}
\caption{Orbital parameters of the simulated interactions. All simulations where run with hyperbolic orbits and for three different velocities 500, 1000, 2000~km/sec}
\label{tab:main}
\end{table}

\section{Effects of the interaction on bar parameters}
\label{sec:effects}

In this section we describe the main results obtained from the simulations. As we stated above, we focus on interactions with pericenters at $30$~kpc. We have also studied those with $15$~kpc, but consider the end product not to be as reliable in this case; we would be using the impulse approximation for an interaction where the disks of both galaxies are overlapping.

We approach the effect of the interaction by analysing some of the main parameters used in the literature to classify and describe stellar bars and their evolution. These are the bar amplitude, bar length, pattern speed and the rotation parameter, ${\cal R}=R_{CR}/R_{bar}$, related to how fast or slow the bar is, and to the mass distribution of the different components, mainly the halo and the disc \citep[e.g.][]{Aguerri+03,Perez+12,Aguerri+15}. 

The bar amplitude is computed as the amplitude of the second coefficient of Fourier density decomposition. In detail, this can be computed by: 

\begin{equation}
A_{m, r} (R) = \frac {1}{N_s} |\sum_j^{N_s} e^{im \theta_j}|, ~~~~~~~m=1, 2, .....
\label{eq:Amr}
\end{equation}

\noindent
where $\theta_j$ is the azimuthal angle of particle $j$ and $N_s$ is the number of particles included in the summation. Here, the summation is over all particles with a cylindrical radius within a range around a given $R$ value, giving a measure of the bar strength at that radius. From the same density distribution, the bar pattern speed is calculated from the phase angle of the bar. 

The bar length is taken at the radius where the ellipticity is $10\%$ lower than the maximum \citep{Martinez-Valpuesta+06}. Finally, $R_{CR}$ is the radius where the angular velocity equals the pattern speed.

\subsection{SIMI1: bar in isolation}

\begin{figure*}
\begin{center}
\includegraphics[scale=0.35]{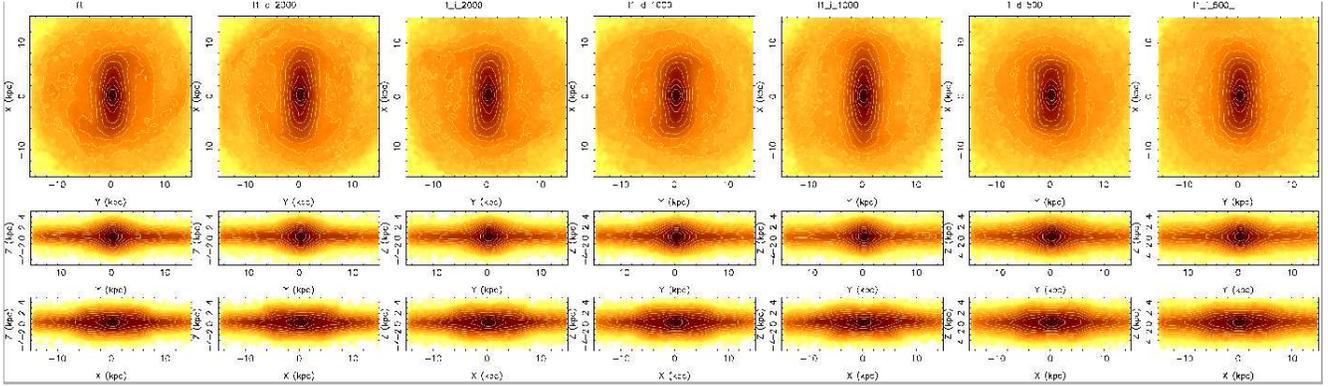}
\caption{Density maps in three projections. The original simulation SIMI1 is plotted on the left, and then from fast to slow interaction. These are plotted at time $\tau=4.18$~Gyr, sometime after the big drop in bar strength to give the bar time to resume its evolution.}
\label{fig:framesI1_11}
\end{center}
\end{figure*}

\begin{figure*}
\begin{center}
\includegraphics[scale=0.6,angle=-90.]{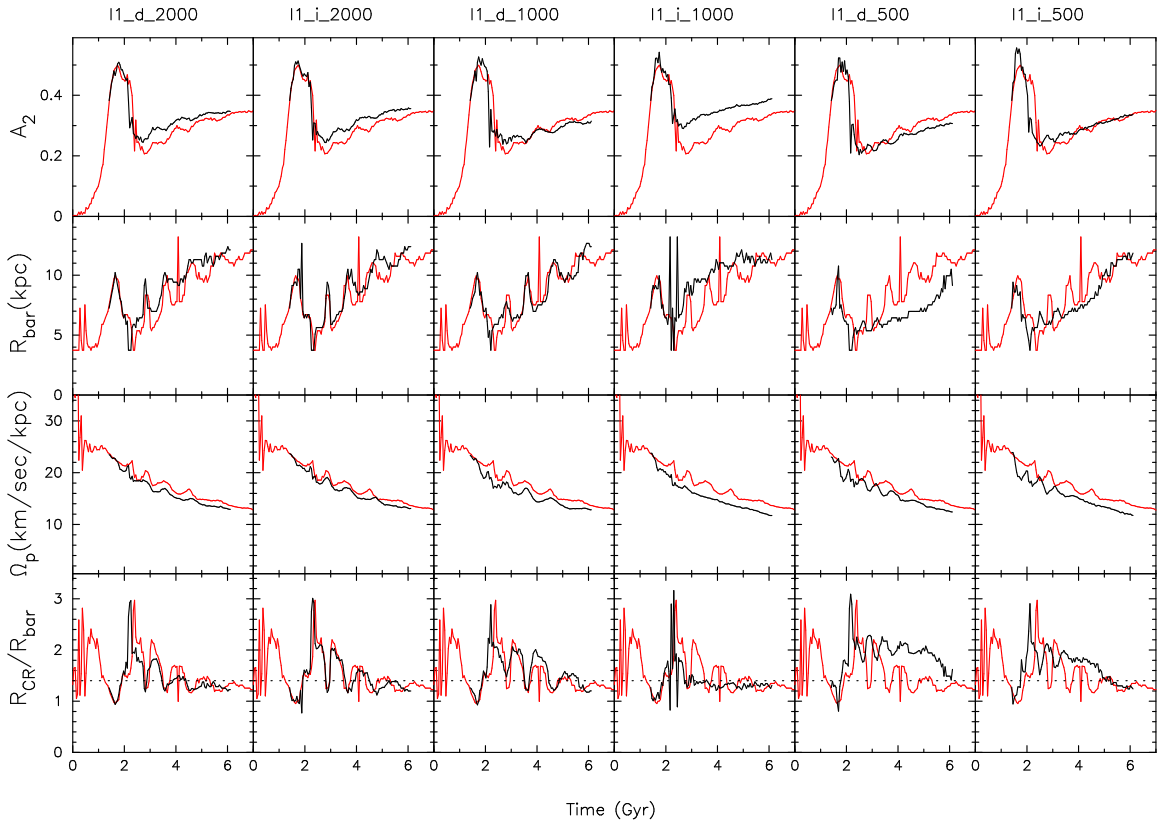}
\caption{Time evolution of bar parameters for the model in isolation, SIMI1 ({\it red line}) and for the simulations of interactions introduced at $t11$. From top to bottom we show bar amplitude, bar length, bar pattern speed and ${\cal R}$.}
\label{fig:I1_11}
\end{center}
\end{figure*}

Simulation SIMI1 was already presented in \cite{Martinez-Valpuesta+Gerhard11}, were it was scaled to match the Milky Way bar length and solar velocity. Here with a different scaling, the disc develops a flat bar in less than 2~Gyr, and by 3~Gyr it already buckles. Then the bar weakens and resumes its evolution, changing its structure and slowly becoming longer, stronger and thicker, forming a peanut/boxy structure when seen edge-on. In \cite{Martinez-Valpuesta+06} there is a more detailed explanation of general bar dynamical and secular evolution, including details of the different buckling events. As a representative snapshot, we choose time $\tau=4.18$~Gyr (Fig.~\ref{fig:framesI1_11}, first column). The three panels show face-on, end-on and side-on projection of SIMI1. The general time evolution of the bar parameters for SIMI1 can be seen in Fig.~\ref{fig:I1_11} (red line). We can see the amplitude of the bar increasing rapidly until the buckling event, then increasing slowly (secular phase). The size of the bar suffers similar evolution. The pattern speed decreases continuously, anticorrelating with the bar amplitude \citep[e.g.,][]{Athanassoula03}. The parameter ${\cal R}$ fluctuates accordingly to the variations of $R_{bar}$ but in general most of the time the bar is fast (${\cal R}\lesssim1.4$).

Our first experiment is based in a 1:1 fly-by with a perturbed galaxy (SIMI1) which has already been able to form and evolve a bar. In order to quantify these characteristics and its effects, we measure the bar strength, the bar length, the pattern speed and ${\cal R}$. In Fig.~\ref{fig:I1_11} we show the evolution of these measurements with time. The resulting bar after the interaction at early and late times ($t11$ and $t12$) has very similar properties to those of the bar created in isolation. The most significant difference, for most of the evolution, is the bar becoming slower in terms of the pattern speed. This slowdown is even bigger when the maximum of the interaction occurs after the bar has already buckled ($t12$). As expected, the effect is always stronger when the interaction is slower (I1\_i\_500). \citet{Gerin+90} has shown that the angular frequency of the bar is not imposed by the perturber and no difference was found between the isolated and the perturbed experiment. This slowdown in angular frequency in our simulations is translated into a slowdown in terms of ${\cal R}$. After the bar buckles and regrows again, the bar becomes considerably shorter, and slower, and for at least $4$~Gyr stays as slow as ${\cal R}>1.8$. In the slow interactions, I1\_i\_500 and  I1\_d\_500, this effect is very important, being up to $\sim30\%$ slower in ${\cal R}$ than in the isolated case.

\subsection{SIMI0: no-bar in isolation}

This simulation, as mentioned previously in Sect.~\ref{sec:Simu}, was conceived to not form a bar in isolation. The main reason is the ratio of dark matter halo mass and disc mass within the inner $7$~kpc. From previous theoretical works \citep{Athanassoula03}, and our own experience, we know that massive central concentration of mass in the inner part at the initial development of the bar ($70\%$ of mass in the halo), and absence of mass to acquire angular momentum in the outer parts, at later times, are the main causes for non developing a bar for more than $6$~Gyr.

\begin{figure*}
\begin{center}
\includegraphics[scale=0.35]{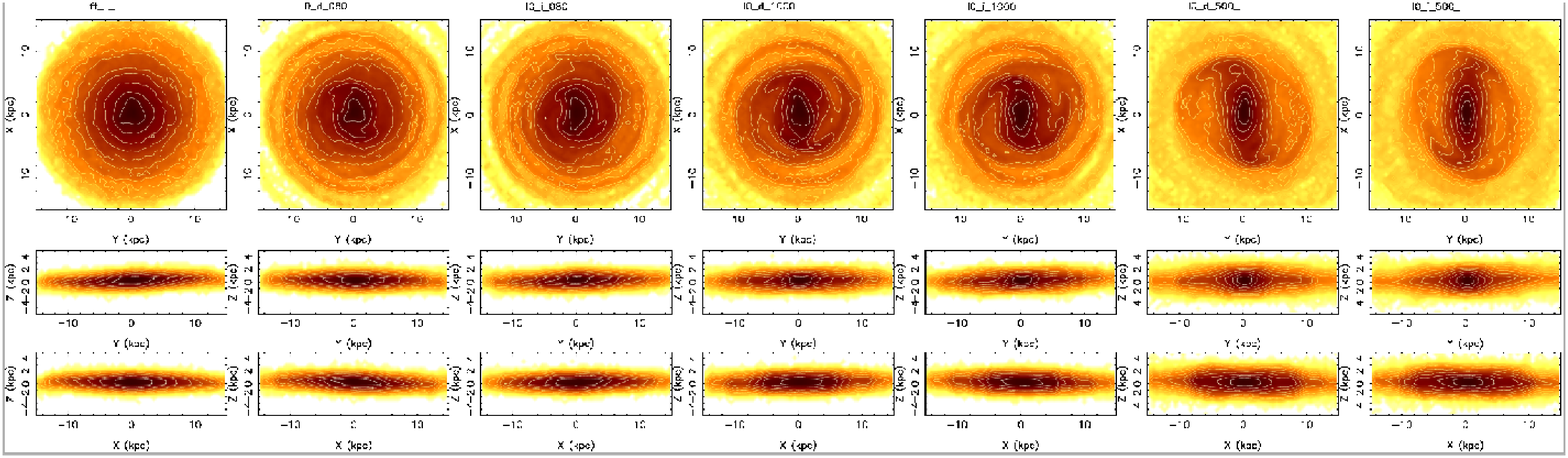}
\caption{Density maps in three projections. The original simulation SIMI0 is plotted on the left, and then from fast to slow interaction. These are plotted time corresponds to time $\tau=4.18$~Gyr}
\label{fig:framesI0_01}
\end{center}
\end{figure*}

\begin{figure*}
\begin{center}
\includegraphics[scale=0.6,angle=-90.]{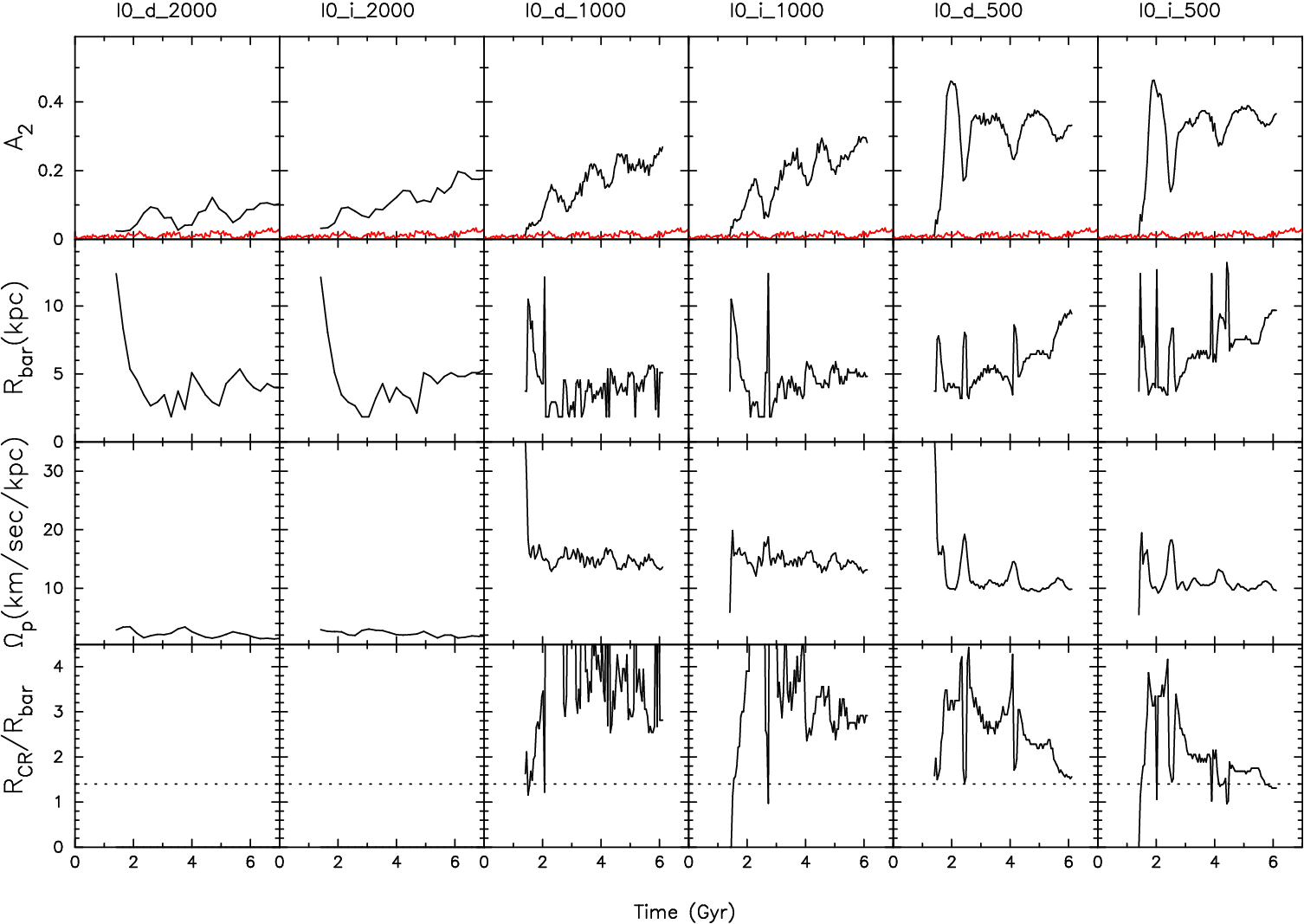}
\caption{Time evolution of bar parameters for the model in isolation, SIMI0 ({\it red line}) and for the simulations of interactions introduced at $t11$. From top to bottom we show bar amplitude, bar length, bar pattern speed and ${\cal R}$.}
\label{fig:I0_01}
\end{center}
\end{figure*}

The interaction by impulse approximation is introduced at time $\tau=1.5$~Gyr. At the time of the interaction, the halo and the disc loose angular momentum because of the tidal distortion. Later on, when the bar is forming, the disc loses angular momentum and the halo gains it, as in the standard angular momentum transfer scenario \citep{ Debattista+98,Debattista+00,Athanassoula03,Martinez-Valpuesta+06}. As we showed before, the slower the interaction the stronger the effect. For our first case, with $v=2000$~km~s$^{-1}$, the effect is small but still noticeable. There is a weak bar developing (Fig.~\ref{fig:framesI0_01}). We can also see how the bar becomes stronger when the velocity of the encounter decreases. Another clear feature of the galaxies affected by the interaction is the amount of structure developed in the disc, from spiral arms to rings (see Fig.~\ref{fig:framesI0_01}).

Let us now describe in detail the evolution of I0\_i\_500. The big amplitude of the $A_2$ mode is initially due to the spiral arms induced by the interactions, and then the strength is due to the recently created bar. The first big drop in bar amplitude at $\tau\sim 2.3$~Gyr (\ref{fig:I0_01} top left panel) of the bar corresponds to a first buckling event. In the second weakening the bar becomes rounder and therefore weaker. The bar length grows together with amplitude. And contrary to standard bar evolution, the pattern speed trend does not anticorrelate with amplitude and stays almost constant with time. Since the bar keeps on growing and the pattern speed is close to constant, the ${\cal R}$ parameter decreases with time. As we show before for the interaction for SIMI1, most of the time the bar is in the slow regime (${\cal R}>1.4$). Towards the end of the simulation, $\tau \sim6$~Gyr, the bar becomes fast with ${\cal R}\sim1.4$.  

In this set of simulations (SIMI0), although some of them develop a bar, the standard bar evolution in simulations is not seen. For example, buckling event and secular growth are seen just in the slow interaction (strong). In the intermediate interaction ($1000$~km~s$^{-1}$, I0\_d\_1000, I0\_i\_1000) the bar is weak, with $A_2\sim0.2$, and keeps on growing in length and strength but does not buckle. In this particular simulation, the growth of the bar in length and the constancy of the pattern speed, $\Omega_p$, results in the bar being very slow with ${\cal R}\sim2$.

\section{Effects of the interaction on photometrical properties}
\label{sec:phot}

Photometric signatures are in general observationally cheaper than kinematic measurements. Therefore we start describing the photometrical signatures for the different bars. {Firstly, we focus on the ellipticity for the set of simulations for SIMI1. Broadly speaking the ellipticity of the bars varies from $0.7$ before the buckling to $\epsilon \simeq 0.5$ after the buckling event and then slowly increases up to $0.6$. The simulations where the ellipticity is smallest are I1\_500, but note that the difference within this set is no more than $\delta \epsilon \leq 0.1$. For the set of simulations SIMI0 the ellipticity in the strongest case, I0\_500, is on average $\sim 0.65$.

In general we have noticed that certain bars influenced by the interaction are clearly more boxy, in face on view, than those created in isolation (Fig.~\ref{fig:framesI1_11}). We have analysed the boxiness parameter, $100(a_4/a)$, which is a clear indicator of boxy isophotes if negative and disky isophotes if positive \citep{Bender88b}. We obtain negative values in the outer parts of the bars, with average value for the I1 series of $100(a4/a)=-0.2$. At the end of the bar the value reaches the minimum. The differences between those bars affected by the interactions are very small, $100(a4/a)\sim 0.1$. For the slow prograde interaction the boxiness parameter reaches $100(a4/a)=-0.25$ for $20\%$ of the bar length. The set SIMI0 is not boxy at all and this is also reflected in a slightly higher ellipticity of the induced bar.

It is clear when looking at density maps (Fig.~\ref{fig:framesI1_11}), that the bar structure has slightly changed in the 3D-view. For example, in the fast interaction (2000~km~s$^{-1}$), both prograde and retrograde, the bar has become longer, and somehow ``more pinched" in the boxy bulge. For the very slow interaction (500~km~s$^{-1}$) the bar becomes shorter and also pinched in the boxy bulge.

In general, bars created in isolation show a boxy-peanut region which extends up to $2/3$~of the bar length \citep{Athanassoula+Misiriotis02}. In our study, the bars induced purely by the interactions show the thick vertical part reaching almost the extension of the whole bar. We also see how clearly the discs hosting bars triggered by the interaction, have a higher number of structures, including rings and long-lived multi-spirals, as clearly seen in Fig.~\ref{fig:framesI0_01}. As a result, the bar and the spiral arms merge sometimes making the bar stronger during those periods. The spiral arms then form a ring and the bar weakens considerably. These features are also related to how kinematically hot the different resulting discs are, as we show in the next section. 

\section{Effects of the interaction on kinematic properties}
\label{sec:kin}

We have seen that bars created or affected by interactions are slower for longer time (always in terms of ${\cal R}>1.4$),  than those created in isolation. In this section we explore the kinematic imprints of this influence. We will compare each of the isolated galaxies, SIMI0 and SIMI1, with those in the corresponding series where the effect of the interaction is stronger, I0\_i\_500 and I1\_i\_500 respectively. In Fig.~\ref{fig:I1_kin} and Fig.~\ref{fig:I0_kin} we show kinematic maps. The aim of these figures is to show general patterns found in those bars affected or created by interactions. 

\subsection{SIMI1 and I1\_i\_500}

\begin{figure*}
\begin{center}
\includegraphics[scale=0.7]{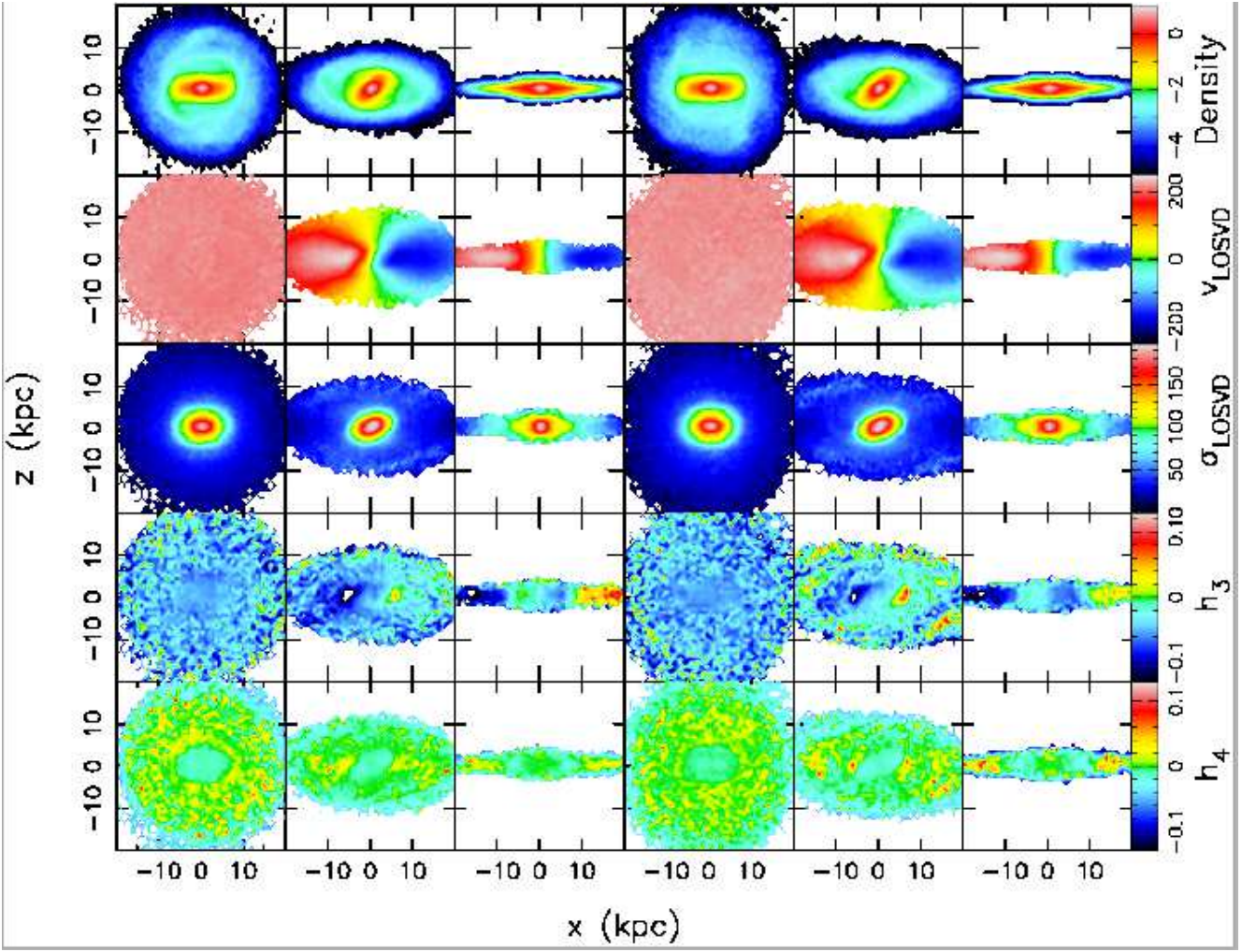}
\caption{From top to bottom density and kinematic moments maps. From left to right the different simulations, SIMI1 simulation at $inc=0^{\circ}$ and $PA_{bar}=0^{\circ}$, $inc=60^{\circ}$ and $PA_{bar}=60^{\circ}$, and $inc=90^{\circ}$ and $PA_{bar}=0^{\circ}$ at $\tau=4.18$~Gyr and for the same orientations for the interaction I1\_i\_500.}
\label{fig:I1_kin}
\end{center}
\end{figure*}

In terms of density distribution in any of the shown views (face-on, mildly inclined and side-on), we cannot see major differences between bars influenced by an interaction and those in isolation (Fig.~\ref{fig:I1_kin}, top row). The same is true for velocity maps. We start seeing the effect of the interaction in the velocity dispersion of the disc, although not in the vertical direction $\sigma_z$. We can clearly see an effect in the mildly inclined frame, which shows a composition of the three velocity components. On average, in the region $R_{bar}<r<1.5R_{bar}$, and on the intermediate inclination ($inc=60^{\circ}$), the velocity dispersion of the disc doubles, from $15$~km~s$^{-1}$ in isolation to $28.5$~km~s$^{-1}$ with the interaction.
But when we look at the edge-on view, we can see the increment in velocity dispersion much better. This is because the interaction increases the radial and the tangential composite velocity dispersion. In the outer parts, for $x>10$~kpc, the value for the isolated case corresponds to $\sim21$~km~s$^{-1}$ and for the interaction to $\sim42$~km~s$^{-1}$. 

For high velocity moments in the Gauss-Hermite decomposition, such as $h_3$ and $h_4$, we see more enhanced structure than for velocity and velocity dispersion. The spiral structure is clearly seen in $h_3$ for the interacting case, in the mildly inclined orientation (Fig.~\ref{fig:I1_kin}, 5th column). The surroundings of the bar also have higher absolute values for both $h_3$ and $h_4$ in comparison with the isolated case. The $v-h_3$ anticorrelation in this case is a clear indicator of disc kinematics. 
As already known the opposite is true for barred kinematics, where the correlation is expected. For example, in the side-on view, in the boxy dominated area, there is a clear correlation $v-h_3$ (for more information on these issues see \citet{Bureau+Athanassoula05, Iannuzzi+Athanassoula15})

\subsection{SIMI0 and I0\_i\_500}

\begin{figure*}
\begin{center}
\includegraphics[scale=0.7]{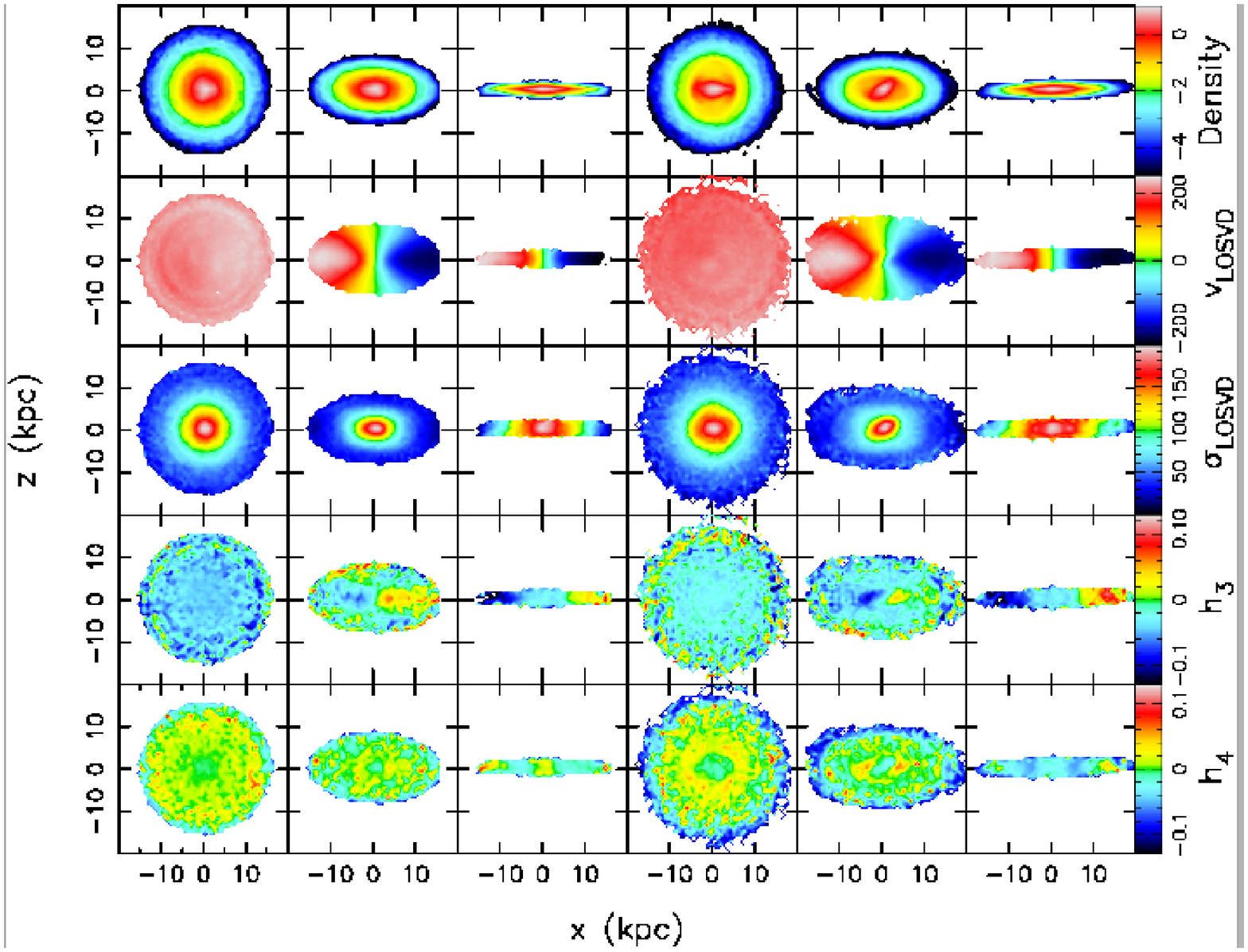}
\caption{From top to bottom density and kinematic moments maps. From left to right the different simulations, SIMI0 simulation at $inc=0^{\circ}$ and $PA_{bar}=0^{\circ}$, $inc=60^{\circ}$ and $PA_{bar}=60^{\circ}$, and $inc=90^{\circ}$ and $PA_{bar}=0^{\circ}$ at $\tau=4.18$~Gyr and for the same orientations for the interaction I0\_i\_500.}
\label{fig:I0_kin}
\end{center}
\end{figure*}

The general characteristics of the density and kinematics moments maps for SIMI0 are standard for a normal non-barred disc galaxy. As before, in the perturbed galaxy we identify more structure in the density map, such as spirals and rings outside the bar (Fig.~\ref{fig:I0_kin}, top row). In the velocity map for the inclined snapshot, we can see the strong twist of the kinematic axis due to the non-circular motions of the bar, and the flares due to the spirals. On average, in the region between $R_{bar}<r<1.5R_{bar}$ the velocity dispersion of the disc is higher by $\sim20$~km~s$^{-1}$, from $3.6$~km~s$^{-1}$ in isolation to $23.9$~km~s$^{-1}$ with the interaction ($inc=60^{\circ}$). In the edge-on view, the vertical parallel isovelocities are clearly associated to the cylindrical rotation of the bar's boxy part. In the dispersion map, as before, the influence of the interaction is clearly seen in the high values of dispersion outside the bar. In the outer parts, for $x>10$~kpc, the value in isolation corresponds to $\sim10$~km~s$^{-1}$ and to $\sim42$~km~s$^{-1}$ for the galaxy in interaction.

We would like to know if the heating comes directly from the bar itself, or from the interaction. When the bar is much weaker due to the very fast interaction (I0\_i\_2000) we do not have such high dispersion in the disc. But at this point we cannot distinguish whether the newly created bar  or the interaction is responsible for the heating.  

For the high velocity moment $h_3$, when the bar is seen edge-on, we should see correlation between $h_3$ and velocity. But in this particular case, I0\_i\_500, where the bar is created purely by interaction, we do not see it as clearly as before. The same is true, for the inclined map (Fig.~\ref{fig:I0_kin}, 5th column), where high values for $h_3$ outside the bar are not so visible.
For $h_4$ maps, in the face-on (Fig.~\ref{fig:I0_kin}, 4th column,bottom) the bar region shows high values, in negative, with the bar shape clearly outlined. Also a ring shape can be seen in $h_4$ around the bar. 

\section{Discussion}
\label{sec:dis}

\subsection{Results and comparison with observations}

In early observational studies of interacting pairs and bars \citep{Elmegreen+90}, the bars were clearly triggered by interactions, in particular those of early type galaxies. This is in good agreement with some of our results, where interactions clearly trigger bars. Conversely, recent statistical analysis of observations \citep{Casteels+13} suggests that bars are suppressed by close interactions between galaxies of similar masses.

In the detailed observational study of galaxy pairs by \citet{CoutodaSilva+deSouza06}, they found no significant change in bar ellipticity with pair separation. Their interpretation of this result was that bar ellipticity is probably governed by intrinsic factors such as velocity field, bulge/disc mass ratio, or mass distribution. This is in complete agreement with our results, since the bar ellipticity evolves depending on initial conditions and in both simulations sets (SIMI1, SIMI0) we found an extended range of ellipticities from $\epsilon\sim 0.5$ to $0.7$.

\citet{Barazza+09} compared bars in fields and clusters, finding that bars of cluster galaxies tend to be slightly longer than those of field galaxies. Their sample had 925 galaxies at $z=0.4-0.8$. Based on our study, this result could be explained as the small effect that a fast interaction has over the galaxy's intrinsic fate. In our case we find, in general, longer bars for those interactions with $2000$~km~s$^{-1}$ (associated to clusters), where the bar growth is almost unaffected by the interaction. This effect is also similar and compatible with that one found by \citet{Li+09}, where red bars (evolved) present higher ellipticity in clustered galaxies.

For at least a decade the fate of bars has been related to intrinsic properties; to the mass ratios of the disc and dark matter halo, and to the central mass concentration \citep{Athanassoula03}. Our study emphasises the need to recognise that a significant fraction of bars will have been triggered only by interactions, independently of the intrinsic properties of the hosting system.

\subsection{Robustness of impulsive approximation}

By using the impulse approximation we are only modelling the influence of the interaction at one particular time. We have checked whether an interaction lasting for a longer time would still give the same results, by running a simulation where the perturber is a set of particles representing the companion. The `companion' follows the orbit in such a way that the maximum of the interaction happens at the same time the previous experiments (with IA). The orbit described by the companion is that of simulation I1\_d\_500.
The resulting main parameters of the bar are comparable with those from the impulsive approximation. The case of the bar amplitude and pattern speed can be seen in Fig.~\ref{fig:test1}. The evolution of the bar is clearly similar to that from the IA. This gives the necessary support and independence to our results obtained with the IA. We have explored the $R_{CR}/R_{bar}$ evolution for this particular case and we find that it is higher because the bar is shorter. So, if anything, the IA underestimates the extent to which the bar strength can be reduced by the interaction.

We have performed the same analysis with a simulation very stable to bar formation with a hotter disk $Q=3$. The impulse approximation is not able to trigger the formation of the bar. With a long interaction performed in the same way as that described above, the galaxy is also not able to develop a bar.

\begin{figure}
\begin{center}
\includegraphics[scale=0.5,angle=-90.]{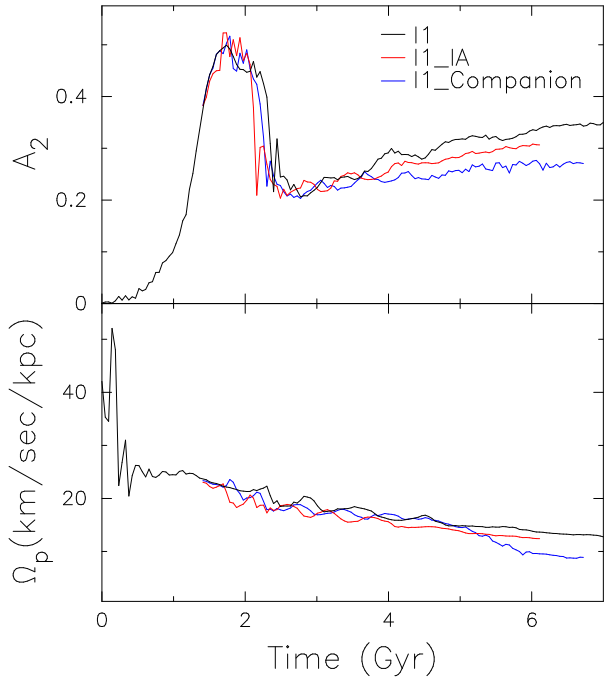}
\caption{Time evolution of bar amplitude and pattern speed for the standard model in isolation ({\it black line}), the interacting model with impulse approximation ({\it red line}) and for the model interacting with a mock galaxy simulating a more lasting interaction ({\it blue line})}
\label{fig:test1}
\end{center}
\end{figure}

\subsection{Prograde versus retrograde orbits}

The early study by \citet[][]{Lang+14} shows that interactions produce stronger bars when the orbit of the encounter is prograde. In our study we find indications of the opposite although the difference between prograde and retrograde is quite small. For example in case I0\_d\_500 the final bar is weaker than in case I0\_i\_500. For a clear effect, although for the case of simulations unstable to bar formation, see I1\_d\_1000 and I1\_i\_1000. In this particular case, there are several effects to be considered. Firstly, if at the initial time the prograde orbit makes the bar stronger, the buckling event will be stronger, and therefore the recovering and the later growth is somehow weaker. We have run the same interaction but at a different time ($t12$), when the bar has already resumed its evolution and the boxy bulge is fully formed. In this case the ending strength of the bar is similar for both orbits, prograde and retrograde.

The second possible explanation is the angle between the bar and the companion at the pericenter at the time of the interaction. Even after studying the simulations with the encounter at different time ($t12$), where the bar angle is different to that at $t11$, we cannot currently distinguish between these two possibilities. 

\subsection{Detecting slow bars}

These simulations are scaled to be compared with galaxies with masses of few times $10^{11}$~$M_{\sun}$. According to the COSMOS survey \citep{Sheth+08}, half of the galaxies in this mass range formed bars at $0.60<z<0.84$, much earlier than lower mass galaxies where just $20\%$ hosted bars at these redshifts. 
Most of the observational studies of pattern speeds of bars in nearby galaxies find most of the bars to be fast. This is in agreement with our study where, some time after the interactions, bars seem to end up in the fast regime with ${\cal R}\leq1.4$.

In the case of massive barred galaxies, we expect that bars created by interactions should be slow in terms of $\cal R$ at intermediate redshift. Conversely, the first and only study of this found that bars are fast at $z\lesssim0.5$ \citep{Perez+12}. This study is purely based in photometry, the survey is not complete in any sense, and the selection is based on galaxies having outer rings, with the possible bias that this can induce.

A way to test our scenario (slow bars being those influenced by interactions) would be to look for pattern speeds of low mass galaxies, since they are assumed to form their bar recently. We could then observe them within the $\sim 4$~Gyr slow-mode time interval that we have predicted with our fly-by simulations. At the moment the only low mass galaxy with a measured pattern speed is NGC~4431 \citep{Corsini+07}. The authors found that the probability of the bar being fast is twice that of being slow. Their study is based on long slit data of just one galaxy, so we should take it with caution. We expect that this galaxy can be studied with 3D kinematic maps in more detail, in particular by looking for signs of interactions in the kinematics. 

\section{Conclusions}
\label{sec:con}

In the last decades, bar formation has almost always been considered as having two causes: instability in isolated discs (self-generated) or triggered by interactions. In this work we run a set of N-body numerical simulations of coplanar 1:1 interactions and explore the differences between these two mechanisms, showing clear differences between them.

\begin{itemize}
\item For galaxies which would form a strong bar in isolation, the interaction was not able to prevent it. The interaction is also not able to strongly change the general evolution of bar parameters. 
\item Conversely, for galaxies which would not form a bar in isolation, a slow interaction developed a strong bar in the galaxy.
\item Bars that were fully triggered or affected by interactions were slower than those created intrinsically by pure dynamical instabilities, and stayed in the slow regime for $4$~Gyr after the closest point of the encounter.
\item As these triggered or affected bars do ultimately speed up, to catch them in the slow phase we should look for them either at high redshifts or in low mass galaxies (where observations indicate that bar formation occurs later).
\item Slow fly-bys, or stronger ones, had a greater effect on the galaxies. Therefore, we expect to find slower bars in low mass groups where the velocity dispersion is lower. 
\item We do not find any consistent differences between prograde or retrograde orbits. 
\item The bar triggered purely by the slow fly-by developed a more radially extended boxy/peanut bulge than any of the isolated simulations. 
\item The effect of fly-bys on the discs as whole was always to kinematically heat them. This was particularly noticeable in the inclined systems. 
\item In those bars triggered by the fly-bys, their discs show more structures such as spirals and rings. If this effect is shown in pure N-body simulations, we expect to effect in real galaxies, with gas and dust, to be even more pronounced.
\end{itemize}

In the future, we will extend our study by using more realistic simulations of full clusters and groups including gas, star formation and feedback. We expect to confirm the results presented here and constrain better the bar and halo properties as well as the disc heating by the interactions.

\section*{Acknowledgements}
We thank Victor Debattista for discussions and suggestions on this work. IMV, MS and CDV thank support from the MINECO through grants AYA2013-46886-P and AYA2014-583308-P and acknowledge financial support from the Spanish Ministry of Economy and Competitiveness (MINECO) under the 2011 Severo Ochoa Program MINECO SEV-2011-0187, and JALA through grant AYA2013-43188-P. GADGET.v2 has been used at early stages of this work (http://www.gadgetcode.org).

\bibliographystyle{mn2e}
%\bibliography{../interactions_5_2}

\begin{thebibliography}{}

\bibitem[\protect\citeauthoryear{{Aguerri}}{{Aguerri}}{1999}]{Aguerri99}
{Aguerri} J.~A.~L.,  1999, \aap, 351, 43

\bibitem[\protect\citeauthoryear{{Aguerri}, {Debattista} \&
  {Corsini}}{{Aguerri} et~al.}{2003}]{Aguerri+03}
{Aguerri} J.~A.~L.,  {Debattista} V.~P.,    {Corsini} E.~M.,  2003, \mnras,
  338, 465

\bibitem[\protect\citeauthoryear{{Aguerri}, {Elias-Rosa}, {Corsini} \&
  {Mu{\~n}oz-Tu{\~n}{\'o}n}}{{Aguerri} et~al.}{2005}]{Aguerri+05}
{Aguerri} J.~A.~L.,  {Elias-Rosa} N.,  {Corsini} E.~M.,
  {Mu{\~n}oz-Tu{\~n}{\'o}n} C.,  2005, \aap, 434, 109

\bibitem[\protect\citeauthoryear{{Aguerri}, {Hunter}, {Prieto}, {Varela},
  {Gottesman} \& {Mu{\~n}oz-Tu{\~n}{\'o}n}}{{Aguerri}
  et~al.}{2001}]{Aguerri+01}
{Aguerri} J.~A.~L.,  {Hunter} J.~H.,  {Prieto} M.,  {Varela} A.~M.,
  {Gottesman} S.~T.,    {Mu{\~n}oz-Tu{\~n}{\'o}n} C.,  2001, \aap, 373, 786

\bibitem[\protect\citeauthoryear{{Aguerri}, {M{\'e}ndez-Abreu} \&
  {Corsini}}{{Aguerri} et~al.}{2009}]{Aguerri+09}
{Aguerri} J.~A.~L.,  {M{\'e}ndez-Abreu} J.,    {Corsini} E.~M.,  2009, \aap,
  495, 491

\bibitem[\protect\citeauthoryear{{Aguerri}, {M{\'e}ndez-Abreu},
  {Falc{\'o}n-Barroso}, {Amorin}, {Barrera-Ballesteros}, {Cid Fernandes} \& {et
  al.}}{{Aguerri} et~al.}{2015}]{Aguerri+15}
{Aguerri} J.~A.~L.,  {M{\'e}ndez-Abreu} J.,  {Falc{\'o}n-Barroso} J.,  {Amorin}
  A.,  {Barrera-Ballesteros} J.,  {Cid Fernandes} R.,    {et al.} 2015, \aap,
  576, A102

\bibitem[\protect\citeauthoryear{{Aguerri}, {Mu{\~n}oz-Tu{\~n}{\'o}n}, {Varela}
  \& {Prieto}}{{Aguerri} et~al.}{2000}]{Aguerri+00}
{Aguerri} J.~A.~L.,  {Mu{\~n}oz-Tu{\~n}{\'o}n} C.,  {Varela} A.~M.,    {Prieto}
  M.,  2000, \aap, 361, 841

\bibitem[\protect\citeauthoryear{{Andersen}}{{Andersen}}{1996}]{Andersen+96}
{Andersen} V.,  1996, \aj, 111, 1805

\bibitem[\protect\citeauthoryear{{Athanassoula}}{{Athanassoula}}{1996}]{Athanassoula96}
{Athanassoula} E.,  1996, in {Sandqvist} A.,  {Lindblad} P.~O.,  eds, Lecture
  Notes in Physics, Berlin Springer Verlag Vol.~474 of Lecture Notes in
  Physics, Berlin Springer Verlag, {The Fate of Barred Galaxies in Interacting
  and Merging Systems}.
p.~59

\bibitem[\protect\citeauthoryear{{Athanassoula}}{{Athanassoula}}{2002}]{Athanassoula02}
{Athanassoula} E.,  2002, \apss, 281, 39

\bibitem[\protect\citeauthoryear{{Athanassoula}}{{Athanassoula}}{2003}]{Athanassoula03}
{Athanassoula} E.,  2003, \mnras, 341, 1179

\bibitem[\protect\citeauthoryear{{Athanassoula}, {Machado} \&
  {Rodionov}}{{Athanassoula} et~al.}{2013}]{Athanassoula+13}
{Athanassoula} E.,  {Machado} R.~E.~G.,    {Rodionov} S.~A.,  2013, \mnras,
  429, 1949

\bibitem[\protect\citeauthoryear{{Athanassoula} \& {Misiriotis}}{{Athanassoula}
  \& {Misiriotis}}{2002}]{Athanassoula+Misiriotis02}
{Athanassoula} E.,  {Misiriotis} A.,  2002, \mnras, 330, 35

\bibitem[\protect\citeauthoryear{{Barazza}, {Jablonka}, {Desai}, {Jogee},
  {Arag{\'o}n-Salamanca}, {De Lucia}, {Saglia}, {Halliday}, {Poggianti},
  {Dalcanton} \& {et al.}}{{Barazza} et~al.}{2009}]{Barazza+09}
{Barazza} F.~D.,  {Jablonka} P.,  {Desai} V.,  {Jogee} S.,
  {Arag{\'o}n-Salamanca} A.,  {De Lucia} G.,  {Saglia} R.~P.,  {Halliday} C.,
  {Poggianti} B.~M.,  {Dalcanton} J.~J.,    {et al.} 2009, \aap, 497, 713

\bibitem[\protect\citeauthoryear{{Bender}}{{Bender}}{1988}]{Bender88b}
{Bender} R.,  1988, \aap, 202, L5

\bibitem[\protect\citeauthoryear{{Berentzen}, {Athanassoula}, {Heller} \&
  {Fricke}}{{Berentzen} et~al.}{2003}]{Berentzen+03}
{Berentzen} I.,  {Athanassoula} E.,  {Heller} C.~H.,    {Fricke} K.~J.,  2003,
  \mnras, 341, 343

\bibitem[\protect\citeauthoryear{{Binney} \& {Tremaine}}{{Binney} \&
  {Tremaine}}{1987}]{Binney+Tremaine87}
{Binney} J.,  {Tremaine} S.,  1987, {Galactic dynamics}

\bibitem[\protect\citeauthoryear{{Bureau} \& {Athanassoula}}{{Bureau} \&
  {Athanassoula}}{2005}]{Bureau+Athanassoula05}
{Bureau} M.,  {Athanassoula} E.,  2005, \apj, 626, 159

\bibitem[\protect\citeauthoryear{{Buta} \& {Block}}{{Buta} \&
  {Block}}{2001}]{buta+Block01}
{Buta} R.,  {Block} D.~L.,  2001, \apj, 550, 243

\bibitem[\protect\citeauthoryear{{Buta} \& {Combes}}{{Buta} \&
  {Combes}}{1996}]{Buta+96}
{Buta} R.,  {Combes} F.,  1996, \fcp, 17, 95

\bibitem[\protect\citeauthoryear{{Buta}, {Laurikainen}, {Salo} \&
  {Knapen}}{{Buta} et~al.}{2010}]{Buta+10}
{Buta} R.,  {Laurikainen} E.,  {Salo} H.,    {Knapen} J.~H.,  2010, \apj, 721,
  259

\bibitem[\protect\citeauthoryear{{Casteels}, {Bamford}, {Skibba}, {Masters},
  {Lintott}, {Keel}, {Schawinski}, {Nichol} \& {Smith}}{{Casteels}
  et~al.}{2013}]{Casteels+13}
{Casteels} K.~R.~V.,  {Bamford} S.~P.,  {Skibba} R.~A.,  {Masters} K.~L.,
  {Lintott} C.~J.,  {Keel} W.~C.,  {Schawinski} K.,  {Nichol} R.~C.,    {Smith}
  A.~M.,  2013, \mnras, 429, 1051

\bibitem[\protect\citeauthoryear{{Combes} \& {Sanders}}{{Combes} \&
  {Sanders}}{1981}]{Combes+Sanders81}
{Combes} F.,  {Sanders} R.~H.,  1981, \aap, 96, 164

\bibitem[\protect\citeauthoryear{{Contopoulos}}{{Contopoulos}}{1981}]{Contopoulos81}
{Contopoulos} G.,  1981, \aap, 102, 265

\bibitem[\protect\citeauthoryear{{Corsini}, {Aguerri}, {Debattista},
  {Pizzella}, {Barazza} \& {Jerjen}}{{Corsini} et~al.}{2007}]{Corsini+07}
{Corsini} E.~M.,  {Aguerri} J.~A.~L.,  {Debattista} V.~P.,  {Pizzella} A.,
  {Barazza} F.~D.,    {Jerjen} H.,  2007, \apjl, 659, L121

\bibitem[\protect\citeauthoryear{{Corsini}, {Debattista} \&
  {Aguerri}}{{Corsini} et~al.}{2003}]{Corsini+03}
{Corsini} E.~M.,  {Debattista} V.~P.,    {Aguerri} J.~A.~L.,  2003, \apjl, 599,
  L29

\bibitem[\protect\citeauthoryear{{Couto da Silva} \& {de Souza}}{{Couto da
  Silva} \& {de Souza}}{2006}]{CoutodaSilva+deSouza06}
{Couto da Silva} T.~C.,  {de Souza} R.~E.,  2006, \aap, 457, 405

\bibitem[\protect\citeauthoryear{{Debattista}, {Corsini} \&
  {Aguerri}}{{Debattista} et~al.}{2002}]{Debattista+02}
{Debattista} V.~P.,  {Corsini} E.~M.,    {Aguerri} J.~A.~L.,  2002, \mnras,
  332, 65

\bibitem[\protect\citeauthoryear{{Debattista} \& {Sellwood}}{{Debattista} \&
  {Sellwood}}{1998}]{Debattista+98}
{Debattista} V.~P.,  {Sellwood} J.~A.,  1998, \apjl, 493, L5

\bibitem[\protect\citeauthoryear{{Debattista} \& {Sellwood}}{{Debattista} \&
  {Sellwood}}{2000}]{Debattista+00}
{Debattista} V.~P.,  {Sellwood} J.~A.,  2000, \apj, 543, 704

\bibitem[\protect\citeauthoryear{{Debattista} \& {Williams}}{{Debattista} \&
  {Williams}}{2004}]{Debattista+04}
{Debattista} V.~P.,  {Williams} T.~B.,  2004, \apj, 605, 714

\bibitem[\protect\citeauthoryear{{Dehnen}}{{Dehnen}}{2002}]{Dehnen02}
{Dehnen} W.,  2002, Journal of Computational Physics, 179, 27

\bibitem[\protect\citeauthoryear{{D{\'{\i}}az-Garc{\'{\i}}a}, {Salo},
  {Laurikainen} \& {Herrera-Endoqui}}{{D{\'{\i}}az-Garc{\'{\i}}a}
  et~al.}{2015}]{Diaz-Garcia+15}
{D{\'{\i}}az-Garc{\'{\i}}a} S.,  {Salo} H.,  {Laurikainen} E.,
  {Herrera-Endoqui} M.,  2015, ArXiv e-prints

% \bibitem[\protect\citeauthoryear{{D'Onghia}, {Vogelsberger}, {Faucher-Giguere}
%   \& {Hernquist}}{{D'Onghia} et~al.}{2010}]{Donghia+10}
% {D'Onghia} E.,  {Vogelsberger} M.,  {Faucher-Giguere} C.-A.,    {Hernquist} L.,
%    2010, \apj, 725, 353

\bibitem[\protect\citeauthoryear{{Dubinski}, {Berentzen} \&
  {Shlosman}}{{Dubinski} et~al.}{2009}]{Dubinski+09}
{Dubinski} J.,  {Berentzen} I.,    {Shlosman} I.,  2009, \apj, 697, 293

\bibitem[\protect\citeauthoryear{{Elmegreen} \& {Elmegreen}}{{Elmegreen} \&
  {Elmegreen}}{1985}]{Elmegreen+95}
{Elmegreen} B.~G.,  {Elmegreen} D.~M.,  1985, \apj, 288, 438

\bibitem[\protect\citeauthoryear{{Elmegreen}, {Elmegreen} \&
  {Bellin}}{{Elmegreen} et~al.}{1990}]{Elmegreen+90}
{Elmegreen} D.~M.,  {Elmegreen} B.~G.,    {Bellin} A.~D.,  1990, \apj, 364, 415

\bibitem[\protect\citeauthoryear{{Erwin}}{{Erwin}}{2005}]{Erwin+05}
{Erwin} P.,  2005, \mnras, 364, 283

\bibitem[\protect\citeauthoryear{{Eskridge}, {Frogel}, {Pogge}, {Quillen},
  {Davies}, {DePoy}, {Houdashelt}, {Kuchinski}, {Ram{\'{\i}}rez}, {Sellgren},
  {Terndrup} \& {Tiede}}{{Eskridge} et~al.}{2000}]{Eskridge+00}
{Eskridge} P.~B.,  {Frogel} J.~A.,  {Pogge} R.~W.,  {Quillen} A.~C.,  {Davies}
  R.~L.,  {DePoy} D.~L.,  {Houdashelt} M.~L.,  {Kuchinski} L.~E.,
  {Ram{\'{\i}}rez} S.~V.,  {Sellgren} K.,  {Terndrup} D.~M.,    {Tiede} G.~P.,
  2000, \aj, 119, 536

\bibitem[\protect\citeauthoryear{{Fall} \& {Efstathiou}}{{Fall} \&
  {Efstathiou}}{1980}]{Fall+Efstathiou80}
{Fall} S.~M.,  {Efstathiou} G.,  1980, \mnras, 193, 189

\bibitem[\protect\citeauthoryear{{Font}, {Beckman}, {Epinat}, {Fathi},
  {Guti{\'e}rrez} \& {Hernandez}}{{Font} et~al.}{2011}]{Font+11}
{Font} J.,  {Beckman} J.~E.,  {Epinat} B.,  {Fathi} K.,  {Guti{\'e}rrez} L.,
  {Hernandez} O.,  2011, \apjl, 741, L14

\bibitem[\protect\citeauthoryear{{Font}, {Beckman}, {Querejeta}, {Epinat},
  {James}, {Blasco-herrera}, {Erroz-Ferrer} \& {P{\'e}rez}}{{Font}
  et~al.}{2014}]{Font+14}
{Font} J.,  {Beckman} J.~E.,  {Querejeta} M.,  {Epinat} B.,  {James} P.~A.,
  {Blasco-herrera} J.,  {Erroz-Ferrer} S.,    {P{\'e}rez} I.,  2014, \apjs,
  210, 2

\bibitem[\protect\citeauthoryear{{Fuentes-Carrera}, {Rosado}, {Amram},
  {Dultzin-Hacyan}, {Cruz-Gonz{\'a}lez}, {Salo}, {Laurikainen}, {Bernal},
  {Ambrocio-Cruz} \& {Le Coarer}}{{Fuentes-Carrera}
  et~al.}{2004}]{Fuentes-Carrera+04}
{Fuentes-Carrera} I.,  {Rosado} M.,  {Amram} P.,  {Dultzin-Hacyan} D.,
  {Cruz-Gonz{\'a}lez} I.,  {Salo} H.,  {Laurikainen} E.,  {Bernal} A.,
  {Ambrocio-Cruz} P.,    {Le Coarer} E.,  2004, \aap, 415, 451

\bibitem[\protect\citeauthoryear{{Gadotti}}{{Gadotti}}{2008}]{Gadotti+08}
{Gadotti} D.~A.,  2008, \mnras, 384, 420

\bibitem[\protect\citeauthoryear{{Gerin}, {Combes} \& {Athanassoula}}{{Gerin}
  et~al.}{1990}]{Gerin+90}
{Gerin} M.,  {Combes} F.,    {Athanassoula} E.,  1990, \aap, 230, 37

\bibitem[\protect\citeauthoryear{{Gerssen}, {Kuijken} \&
  {Merrifield}}{{Gerssen} et~al.}{1999}]{Gerssen+99}
{Gerssen} J.,  {Kuijken} K.,    {Merrifield} M.~R.,  1999, \mnras, 306, 926

\bibitem[\protect\citeauthoryear{{Giuricin}, {Mardirossian}, {Mezzetti} \&
  {Monaco}}{{Giuricin} et~al.}{1993}]{Giuricin+93}
{Giuricin} G.,  {Mardirossian} F.,  {Mezzetti} M.,    {Monaco} P.,  1993, \apj,
  407, 22

\bibitem[\protect\citeauthoryear{{Gnedin}, {Hernquist} \& {Ostriker}}{{Gnedin}
  et~al.}{1999}]{Gnedin+99}
{Gnedin} O.~Y.,  {Hernquist} L.,    {Ostriker} J.~P.,  1999, \apj, 514, 109

\bibitem[\protect\citeauthoryear{{Gonz{\'a}lez-Garc{\'{\i}}a}, {Aguerri} \&
  {Balcells}}{{Gonz{\'a}lez-Garc{\'{\i}}a} et~al.}{2005}]{Gonzalez-Garcia+05}
{Gonz{\'a}lez-Garc{\'{\i}}a} A.~C.,  {Aguerri} J.~A.~L.,    {Balcells} M.,
  2005, \aap, 444, 803

\bibitem[\protect\citeauthoryear{{Gonz{\'a}lez-Garc{\'{\i}}a} \& {van
  Albada}}{{Gonz{\'a}lez-Garc{\'{\i}}a} \& {van
  Albada}}{2005}]{Gonzalez-Garcia+VanAlbada05}
{Gonz{\'a}lez-Garc{\'{\i}}a} A.~C.,  {van Albada} T.~S.,  2005, \mnras, 361,
  1030

\bibitem[\protect\citeauthoryear{{Heller} \& {Shlosman}}{{Heller} \&
  {Shlosman}}{1994}]{Heller+Shlosman94}
{Heller} C.~H.,  {Shlosman} I.,  1994, \apj, 424, 84

\bibitem[\protect\citeauthoryear{{Hernquist} \& {Mihos}}{{Hernquist} \&
  {Mihos}}{1995}]{Hernquist+95}
{Hernquist} L.,  {Mihos} J.~C.,  1995, \apj, 448, 41

\bibitem[\protect\citeauthoryear{{Iannuzzi} \& {Athanassoula}}{{Iannuzzi} \&
  {Athanassoula}}{2015}]{Iannuzzi+Athanassoula15}
{Iannuzzi} F.,  {Athanassoula} E.,  2015, \mnras, 450, 2514

\bibitem[\protect\citeauthoryear{{Kazantzidis}, {{\L}okas}, {Callegari},
  {Mayer} \& {Moustakas}}{{Kazantzidis} et~al.}{2011}]{Kazantzidis+11}
{Kazantzidis} S.,  {{\L}okas} E.~L.,  {Callegari} S.,  {Mayer} L.,
  {Moustakas} L.~A.,  2011, \apj, 726, 98

\bibitem[\protect\citeauthoryear{{Kent}}{{Kent}}{1987}]{Kent+87}
{Kent} S.~M.,  1987, \aj, 93, 1062

\bibitem[\protect\citeauthoryear{{Kormendy}}{{Kormendy}}{1979}]{Kormendy79}
{Kormendy} J.,  1979, \apj, 227, 714

\bibitem[\protect\citeauthoryear{{Laine} \& {Heller}}{{Laine} \&
  {Heller}}{1999}]{Laine+99}
{Laine} S.,  {Heller} C.~H.,  1999, \mnras, 308, 557

\bibitem[\protect\citeauthoryear{{Laine}, {Shlosman}, {Knapen} \&
  {Peletier}}{{Laine} et~al.}{2002}]{Laine+02}
{Laine} S.,  {Shlosman} I.,  {Knapen} J.~H.,    {Peletier} R.~F.,  2002, \apj,
  567, 97

\bibitem[\protect\citeauthoryear{{Lang}, {Holley-Bockelmann} \& {Sinha}}{{Lang}
  et~al.}{2014}]{Lang+14}
{Lang} M.,  {Holley-Bockelmann} K.,    {Sinha} M.,  2014, \apjl, 790, L33

\bibitem[\protect\citeauthoryear{{Lansbury}, {Lucey} \& {Smith}}{{Lansbury}
  et~al.}{2014}]{Lansbury+14}
{Lansbury} G.~B.,  {Lucey} J.~R.,    {Smith} R.~J.,  2014, \mnras, 439, 1749

\bibitem[\protect\citeauthoryear{{Laurikainen}, {Salo} \& {Buta}}{{Laurikainen}
  et~al.}{2005}]{Laurikainen+05}
{Laurikainen} E.,  {Salo} H.,    {Buta} R.,  2005, \mnras, 362, 1319

\bibitem[\protect\citeauthoryear{{Laurikainen}, {Salo}, {Buta} \&
  {Knapen}}{{Laurikainen} et~al.}{2007}]{Laurikainen+07}
{Laurikainen} E.,  {Salo} H.,  {Buta} R.,    {Knapen} J.~H.,  2007, \mnras,
  381, 401

\bibitem[\protect\citeauthoryear{{Laurikainen}, {Salo}, {Buta} \&
  {Knapen}}{{Laurikainen} et~al.}{2009}]{Laurikainen+09}
{Laurikainen} E.,  {Salo} H.,  {Buta} R.,    {Knapen} J.~H.,  2009, \apjl, 692,
  L34

\bibitem[\protect\citeauthoryear{{Li}, {Gadotti}, {Mao} \& {Kauffmann}}{{Li}
  et~al.}{2009}]{Li+09}
{Li} C.,  {Gadotti} D.~A.,  {Mao} S.,    {Kauffmann} G.,  2009, \mnras, 397,
  726

\bibitem[\protect\citeauthoryear{{Lin}, {Cervantes Sodi}, {Li}, {Wang} \&
  {Wang}}{{Lin} et~al.}{2014}]{Lin+14}
{Lin} Y.,  {Cervantes Sodi} B.,  {Li} C.,  {Wang} L.,    {Wang} E.,  2014,
  \apj, 796, 98

\bibitem[\protect\citeauthoryear{{Lindblad}, {Lindblad} \&
  {Athanassoula}}{{Lindblad} et~al.}{1996}]{Lindblad+96}
{Lindblad} P.~A.~B.,  {Lindblad} P.~O.,    {Athanassoula} E.,  1996, \aap, 313,
  65

\bibitem[\protect\citeauthoryear{{{\L}okas}, {Athanassoula}, {Debattista},
  {Valluri}, {Pino}, {Semczuk}, {Gajda} \& {Kowalczyk}}{{{\L}okas}
  et~al.}{2014}]{Lokas+14}
{{\L}okas} E.~L.,  {Athanassoula} E.,  {Debattista} V.~P.,  {Valluri} M.,
  {Pino} A.~d.,  {Semczuk} M.,  {Gajda} G.,    {Kowalczyk} K.,  2014, \mnras,
  445, 1339

\bibitem[\protect\citeauthoryear{{Lynden-Bell} \& {Kalnajs}}{{Lynden-Bell} \&
  {Kalnajs}}{1972}]{Lynden-Bell+72}
{Lynden-Bell} D.,  {Kalnajs} A.~J.,  1972, \mnras, 157, 1

\bibitem[\protect\citeauthoryear{{Marinova} \& {Jogee}}{{Marinova} \&
  {Jogee}}{2007}]{Marinova+07}
{Marinova} I.,  {Jogee} S.,  2007, \apj, 659, 1176

\bibitem[\protect\citeauthoryear{{M{\'a}rquez}, {Durret}, {Gonz{\'a}lez
  Delgado}, {Marrero}, {Masegosa}, {Maza}, {Moles}, {P{\'e}rez} \&
  {Roth}}{{M{\'a}rquez} et~al.}{1999}]{Marquez+99}
{M{\'a}rquez} I.,  {Durret} F.,  {Gonz{\'a}lez Delgado} R.~M.,  {Marrero} I.,
  {Masegosa} J.,  {Maza} J.,  {Moles} M.,  {P{\'e}rez} E.,    {Roth} M.,  1999,
  \aaps, 140, 1

\bibitem[\protect\citeauthoryear{{Marquez}, {Moles} \& {Masegosa}}{{Marquez}
  et~al.}{1996}]{Marquez+96}
{Marquez} I.,  {Moles} M.,    {Masegosa} J.,  1996, \aap, 310, 401

\bibitem[\protect\citeauthoryear{{Martin}}{{Martin}}{1995}]{Martin95}
{Martin} P.,  1995, \aj, 109, 2428

\bibitem[\protect\citeauthoryear{{Martinet} \& {Friedli}}{{Martinet} \&
  {Friedli}}{1997}]{Martinet+Friedli97}
{Martinet} L.,  {Friedli} D.,  1997, \aap, 323, 363

\bibitem[\protect\citeauthoryear{{Martinez-Valpuesta} \&
  {Gerhard}}{{Martinez-Valpuesta} \&
  {Gerhard}}{2011}]{Martinez-Valpuesta+Gerhard11}
{Martinez-Valpuesta} I.,  {Gerhard} O.,  2011, \apjl, 734, L20

\bibitem[\protect\citeauthoryear{{Martinez-Valpuesta} \&
  {Shlosman}}{{Martinez-Valpuesta} \&
  {Shlosman}}{2004}]{Martinez-Valpuesta+Shlosman04}
{Martinez-Valpuesta} I.,  {Shlosman} I.,  2004, \apjl, 613, L29

\bibitem[\protect\citeauthoryear{{Martinez-Valpuesta}, {Shlosman} \&
  {Heller}}{{Martinez-Valpuesta} et~al.}{2006}]{Martinez-Valpuesta+06}
{Martinez-Valpuesta} I.,  {Shlosman} I.,    {Heller} C.,  2006, \apj, 637, 214

\bibitem[\protect\citeauthoryear{{Masters}, {Nichol}, {Hoyle}, {Lintott},
  {Bamford}, {Edmondson}, {Fortson}, {Keel}, {Schawinski}, {Smith} \&
  {Thomas}}{{Masters} et~al.}{2011}]{Masters+11}
{Masters} K.~L.,  {Nichol} R.~C.,  {Hoyle} B.,  {Lintott} C.,  {Bamford} S.~P.,
   {Edmondson} E.~M.,  {Fortson} L.,  {Keel} W.~C.,  {Schawinski} K.,  {Smith}
  A.~M.,    {Thomas} D.,  2011, \mnras, 411, 2026

\bibitem[\protect\citeauthoryear{{Melvin}, {Masters}, {Lintott}, {Nichol},
  {Simmons}, {Bamford}, {Casteels}, {Cheung}, {Edmondson}, {Fortson},
  {Schawinski}, {Skibba}, {Smith} \& {Willett}}{{Melvin}
  et~al.}{2014}]{Melvin+14}
{Melvin} T.,  {Masters} K.,  {Lintott} C.,  {Nichol} R.~C.,  {Simmons} B.,
  {Bamford} S.~P.,  {Casteels} K.~R.~V.,  {Cheung} E.,  {Edmondson} E.~M.,
  {Fortson} L.,  {Schawinski} K.,  {Skibba} R.~A.,  {Smith} A.~M.,    {Willett}
  K.~W.,  2014, \mnras, 438, 2882

\bibitem[\protect\citeauthoryear{{M{\'e}ndez-Abreu}, {S{\'a}nchez-Janssen},
  {Aguerri}, {Corsini} \& {Zarattini}}{{M{\'e}ndez-Abreu}
  et~al.}{2012}]{Mendez-Abreu+12}
{M{\'e}ndez-Abreu} J.,  {S{\'a}nchez-Janssen} R.,  {Aguerri} J.~A.~L.,
  {Corsini} E.~M.,    {Zarattini} S.,  2012, \apjl, 761, L6

\bibitem[\protect\citeauthoryear{{Men{\'e}ndez-Delmestre}, {Sheth},
  {Schinnerer}, {Jarrett} \& {Scoville}}{{Men{\'e}ndez-Delmestre}
  et~al.}{2007}]{Menendez-Delmestre+07}
{Men{\'e}ndez-Delmestre} K.,  {Sheth} K.,  {Schinnerer} E.,  {Jarrett} T.~H.,
   {Scoville} N.~Z.,  2007, \apj, 657, 790

\bibitem[\protect\citeauthoryear{{Merrifield} \& {Kuijken}}{{Merrifield} \&
  {Kuijken}}{1995}]{Merrifield+95}
{Merrifield} M.~R.,  {Kuijken} K.,  1995, \mnras, 274, 933

\bibitem[\protect\citeauthoryear{{Miwa} \& {Noguchi}}{{Miwa} \&
  {Noguchi}}{1998}]{Miwa+Noguchi98}
{Miwa} T.,  {Noguchi} M.,  1998, \apj, 499, 149

\bibitem[\protect\citeauthoryear{{Mu{\~n}oz-Tu{\~n}{\'o}n}, {Caon} \&
  {Aguerri}}{{Mu{\~n}oz-Tu{\~n}{\'o}n} et~al.}{2004}]{Munoz-Tunon+04}
{Mu{\~n}oz-Tu{\~n}{\'o}n} C.,  {Caon} N.,    {Aguerri} J.~A.~L.,  2004, \aj,
  127, 58

\bibitem[\protect\citeauthoryear{{Nair} \& {Abraham}}{{Nair} \&
  {Abraham}}{2010}]{Nair+10}
{Nair} P.~B.,  {Abraham} R.~G.,  2010, \apjl, 714, L260

\bibitem[\protect\citeauthoryear{{Noguchi}}{{Noguchi}}{1987}]{Noguchi87}
{Noguchi} M.,  1987, \mnras, 228, 635

\bibitem[\protect\citeauthoryear{{Ohta}, {Hamabe} \& {Wakamatsu}}{{Ohta}
  et~al.}{1990}]{Ohta+90}
{Ohta} K.,  {Hamabe} M.,    {Wakamatsu} K.-I.,  1990, \apj, 357, 71

\bibitem[\protect\citeauthoryear{{P{\'e}rez}, {Aguerri} \&
  {M{\'e}ndez-Abreu}}{{P{\'e}rez} et~al.}{2012}]{Perez+12}
{P{\'e}rez} I.,  {Aguerri} J.~A.~L.,    {M{\'e}ndez-Abreu} J.,  2012, \aap,
  540, A103

\bibitem[\protect\citeauthoryear{{P{\'e}rez}, {Fux} \& {Freeman}}{{P{\'e}rez}
  et~al.}{2004}]{Perez+04}
{P{\'e}rez} I.,  {Fux} R.,    {Freeman} K.,  2004, \aap, 424, 799

\bibitem[\protect\citeauthoryear{{Prieto}, {Aguerri}, {Varela} \&
  {Mu{\~n}oz-Tu{\~n}{\'o}n}}{{Prieto} et~al.}{2001}]{Prieto+01}
{Prieto} M.,  {Aguerri} J.~A.~L.,  {Varela} A.~M.,    {Mu{\~n}oz-Tu{\~n}{\'o}n}
  C.,  2001, \aap, 367, 405

\bibitem[\protect\citeauthoryear{{Prieto}, {Gottesman}, {Aguerri} \&
  {Varela}}{{Prieto} et~al.}{1997}]{Prieto+97}
{Prieto} M.,  {Gottesman} S.~T.,  {Aguerri} J.-A.~L.,    {Varela} A.-M.,  1997,
  \aj, 114, 1413

\bibitem[\protect\citeauthoryear{{Puerari} \& {Dottori}}{{Puerari} \&
  {Dottori}}{1997}]{Puerari+97}
{Puerari} I.,  {Dottori} H.,  1997, \apjl, 476, L73

\bibitem[\protect\citeauthoryear{{Quillen}, {Frogel} \& {Gonzalez}}{{Quillen}
  et~al.}{1994}]{Quillen+94}
{Quillen} A.~C.,  {Frogel} J.~A.,    {Gonzalez} R.~A.,  1994, \apj, 437, 162

\bibitem[\protect\citeauthoryear{{Raha}, {Sellwood}, {James} \& {Kahn}}{{Raha}
  et~al.}{1991}]{Raha+91}
{Raha} N.,  {Sellwood} J.~A.,  {James} R.~A.,    {Kahn} F.~D.,  1991, \nat,
  352, 411

\bibitem[\protect\citeauthoryear{{Romano-D{\'\i}az}, {Shlosman}, {Heller} \&
  {Hoffman}}{{Romano-D{\'\i}az} et~al.}{2008}]{Romano-Diaz+08}
{Romano-D{\'\i}az} E.,  {Shlosman} I.,  {Heller} C.,    {Hoffman} Y.,  2008,
  \apjl, 687, L13

\bibitem[\protect\citeauthoryear{{Saha}, {Martinez-Valpuesta} \&
  {Gerhard}}{{Saha} et~al.}{2012}]{Saha+12}
{Saha} K.,  {Martinez-Valpuesta} I.,    {Gerhard} O.,  2012, \mnras, 421, 333

\bibitem[\protect\citeauthoryear{{Salo}, {Laurikainen}, {Buta} \&
  {Knapen}}{{Salo} et~al.}{2010}]{Salo+10}
{Salo} H.,  {Laurikainen} E.,  {Buta} R.,    {Knapen} J.~H.,  2010, \apjl, 715,
  L56

\bibitem[\protect\citeauthoryear{{Seidel}, {Falc{\'o}n-Barroso},
  {Mart{\'{\i}}nez-Valpuesta}, {D{\'{\i}}az-Garc{\'{\i}}a}, {Laurikainen},
  {Salo} \& {Knapen}}{{Seidel} et~al.}{2015}]{Seidel+15}
{Seidel} M.~K.,  {Falc{\'o}n-Barroso} J.,  {Mart{\'{\i}}nez-Valpuesta} I.,
  {D{\'{\i}}az-Garc{\'{\i}}a} S.,  {Laurikainen} E.,  {Salo} H.,    {Knapen}
  J.~H.,  2015, \mnras, 451, 936

\bibitem[\protect\citeauthoryear{{Sempere}, {Garcia-Burillo}, {Combes} \&
  {Knapen}}{{Sempere} et~al.}{1995}]{Sempere+95}
{Sempere} M.~J.,  {Garcia-Burillo} S.,  {Combes} F.,    {Knapen} J.~H.,  1995,
  \aap, 296, 45

\bibitem[\protect\citeauthoryear{{Sheth}, {Elmegreen}, {Elmegreen}, {Capak},
  {Abraham}, {Athanassoula}, {Ellis}, {Mobasher}, {Salvato}, {Schinnerer},
  {Scoville}, {Spalsbury}, {Strubbe}, {Carollo}, {Rich} \& {West}}{{Sheth}
  et~al.}{2008}]{Sheth+08}
{Sheth} K.,  {Elmegreen} D.~M.,  {Elmegreen} B.~G.,  {Capak} P.,  {Abraham}
  R.~G.,  {Athanassoula} E.,  {Ellis} R.~S.,  {Mobasher} B.,  {Salvato} M.,
  {Schinnerer} E.,  {Scoville} N.~Z.,  {Spalsbury} L.,  {Strubbe} L.,
  {Carollo} M.,  {Rich} M.,    {West} A.~A.,  2008, \apj, 675, 1141

\bibitem[\protect\citeauthoryear{{Shlosman}, {Begelman} \& {Frank}}{{Shlosman}
  et~al.}{1990}]{Shlosman+90}
{Shlosman} I.,  {Begelman} M.~C.,    {Frank} J.,  1990, \nat, 345, 679

\bibitem[\protect\citeauthoryear{{Sierra}, {Seigar}, {Treuthardt} \&
  {Puerari}}{{Sierra} et~al.}{2014}]{Sierra+14}
{Sierra} A.,  {Seigar} M.~S.,  {Treuthardt} P.,    {Puerari} I.,  2014, in
  {Seigar} M.~S.,  {Treuthardt} P.,  eds, Structure and Dynamics of Disk
  Galaxies Vol.~480 of Astronomical Society of the Pacific Conference Series,
  {Determination of Resonance Locations in NGC 4145 Using Multiband
  Photometry}.
p.~65

\bibitem[\protect\citeauthoryear{{Sundin}, {Donner} \& {Sundelius}}{{Sundin}
  et~al.}{1993}]{Sundin+93}
{Sundin} M.,  {Donner} K.~J.,    {Sundelius} B.,  1993, \aap, 280, 105

\bibitem[\protect\citeauthoryear{{Thompson}}{{Thompson}}{1981}]{Thompson+81}
{Thompson} L.~A.,  1981, \apjl, 244, L43

\bibitem[\protect\citeauthoryear{{Tremaine} \& {Weinberg}}{{Tremaine} \&
  {Weinberg}}{1984}]{Tremaine+84}
{Tremaine} S.,  {Weinberg} M.~D.,  1984, \mnras, 209, 729

\bibitem[\protect\citeauthoryear{{Treuthardt}, {Buta}, {Salo} \&
  {Laurikainen}}{{Treuthardt} et~al.}{2007}]{Treuthardt+07}
{Treuthardt} P.,  {Buta} R.,  {Salo} H.,    {Laurikainen} E.,  2007, \aj, 134,
  1195

\bibitem[\protect\citeauthoryear{{Treuthardt}, {Salo}, {Rautiainen} \&
  {Buta}}{{Treuthardt} et~al.}{2008}]{Treuthardt+08}
{Treuthardt} P.,  {Salo} H.,  {Rautiainen} P.,    {Buta} R.,  2008, \aj, 136,
  300

\bibitem[\protect\citeauthoryear{{Weinberg}}{{Weinberg}}{1985}]{Weinberg+85}
{Weinberg} M.~D.,  1985, \mnras, 213, 451

\bibitem[\protect\citeauthoryear{{Weinberg} \& {Katz}}{{Weinberg} \&
  {Katz}}{2007a}]{Weinberg+Katz07a}
{Weinberg} M.~D.,  {Katz} N.,  2007a, \mnras, 375, 425

\bibitem[\protect\citeauthoryear{{Weinberg} \& {Katz}}{{Weinberg} \&
  {Katz}}{2007b}]{Weinberg+Katz07b}
{Weinberg} M.~D.,  {Katz} N.,  2007b, \mnras, 375, 460

\bibitem[\protect\citeauthoryear{{Weiner}, {Sellwood} \& {Williams}}{{Weiner}
  et~al.}{2001}]{Weiner+01}
{Weiner} B.~J.,  {Sellwood} J.~A.,    {Williams} T.~B.,  2001, \apj, 546, 931

\bibitem[\protect\citeauthoryear{{Weinzirl}, {Jogee}, {Khochfar}, {Burkert} \&
  {Kormendy}}{{Weinzirl} et~al.}{2009}]{Weinzirl+09}
{Weinzirl} T.,  {Jogee} S.,  {Khochfar} S.,  {Burkert} A.,    {Kormendy} J.,
  2009, \apj, 696, 411

\bibitem[\protect\citeauthoryear{{Whyte}, {Abraham}, {Merrifield}, {Eskridge},
  {Frogel} \& {Pogge}}{{Whyte} et~al.}{2002}]{Whyte+02}
{Whyte} L.~F.,  {Abraham} R.~G.,  {Merrifield} M.~R.,  {Eskridge} P.~B.,
  {Frogel} J.~A.,    {Pogge} R.~W.,  2002, \mnras, 336, 1281

\bibitem[\protect\citeauthoryear{{Wozniak}, {Friedli}, {Martinet}, {Martin} \&
  {Bratschi}}{{Wozniak} et~al.}{1995}]{Wozniak+95}
{Wozniak} H.,  {Friedli} D.,  {Martinet} L.,  {Martin} P.,    {Bratschi} P.,
  1995, \aaps, 111, 115

\end{thebibliography}

%===================================================================
\label{lastpage}

\end{document}